\pdfoutput=1

\documentclass[11pt]{article}

\usepackage[]{acl}

\usepackage{times}
\usepackage{latexsym}

\usepackage[T1]{fontenc}
\usepackage{booktabs}
\usepackage{graphicx} 
\usepackage{hyperref}
\usepackage{color}
\usepackage{pythonhighlight}
\usepackage{graphicx}
\usepackage{pgfplots}
\usepackage[utf8]{inputenc}

\usepackage{microtype}
\usepackage{amsmath,lipsum}

\usepackage{inconsolata}
\usepackage{bigfoot}

\usepackage[english]{babel}
\usepackage{blindtext}

\title{Re3val: Reinforced and Reranked Generative Retrieval}

\author{\textbf{EuiYul Song}$^{1\dagger}$~~~\textbf{Sangryul Kim}$^{2}$~~~\textbf{Haeju Lee}$^{3 \dagger}$~~~\textbf{Joonkee Kim}$^{2}$~~~\textbf{James Thorne}$^{2}$ \smallskip \\ $^1${Samsung Electronics} , \texttt{euiyul.song@samsung.com} \\ $^2$KAIST AI , \texttt{\{sangryul,joonkeekim,thorne\}@kaist.ac.kr} \\ $^3$LG AI Research , \texttt{haeju.lee@lgresearch.ai}}

\begin{document}

\maketitle
\begingroup\def\thefootnote{$\dagger$}\footnotetext{Work performed while at KAIST AI.}\endgroup

\begin{abstract}
Generative retrieval models encode pointers to information in a corpus as an index within the model's parameters. These models serve as part of a larger pipeline, where retrieved information conditions generation for knowledge-intensive NLP tasks. However, we identify two limitations: the generative retrieval does not account for contextual information. Secondly, the retrieval can't be tuned for the downstream readers as decoding the page title is a non-differentiable operation. This paper introduces Re3val, trained with generative reranking and reinforcement learning using limited data. Re3val leverages context acquired via Dense Passage Retrieval to rerank the retrieved page titles and utilizes REINFORCE to maximize rewards generated by constrained decoding. Additionally, we generate questions from our pre-training dataset to mitigate epistemic uncertainty and bridge the domain gap between the pre-training and fine-tuning datasets. Subsequently, we extract and rerank contexts from the KILT database using the rerank page titles. Upon grounding the top five reranked contexts, Re3val demonstrates the Top 1 KILT scores compared to all other generative retrieval models across five KILT datasets. 






\end{abstract}

\section{Introduction}
The primary objective of retrieval models is to enhance the accuracy of answers by selecting the most relevant documents retrieved for a given query, ensuring models have sufficient information to help the downstream reasoning process. For instance, DRQA \citep{chen-etal-2017-reading} introduces a "retrieve and read" pipeline using TF-IDF to return documents for a question answering model to achieve this goal. More recently, NLP researchers have studied neural retrieval models like Dense Passage Retrieval (DPR) \citep{karpukhin-etal-2020-dense} with a seq2seq model to build retrieval augmented language models.
\begin{figure}[t]
    \centering
    \includegraphics[width=1.0\linewidth]{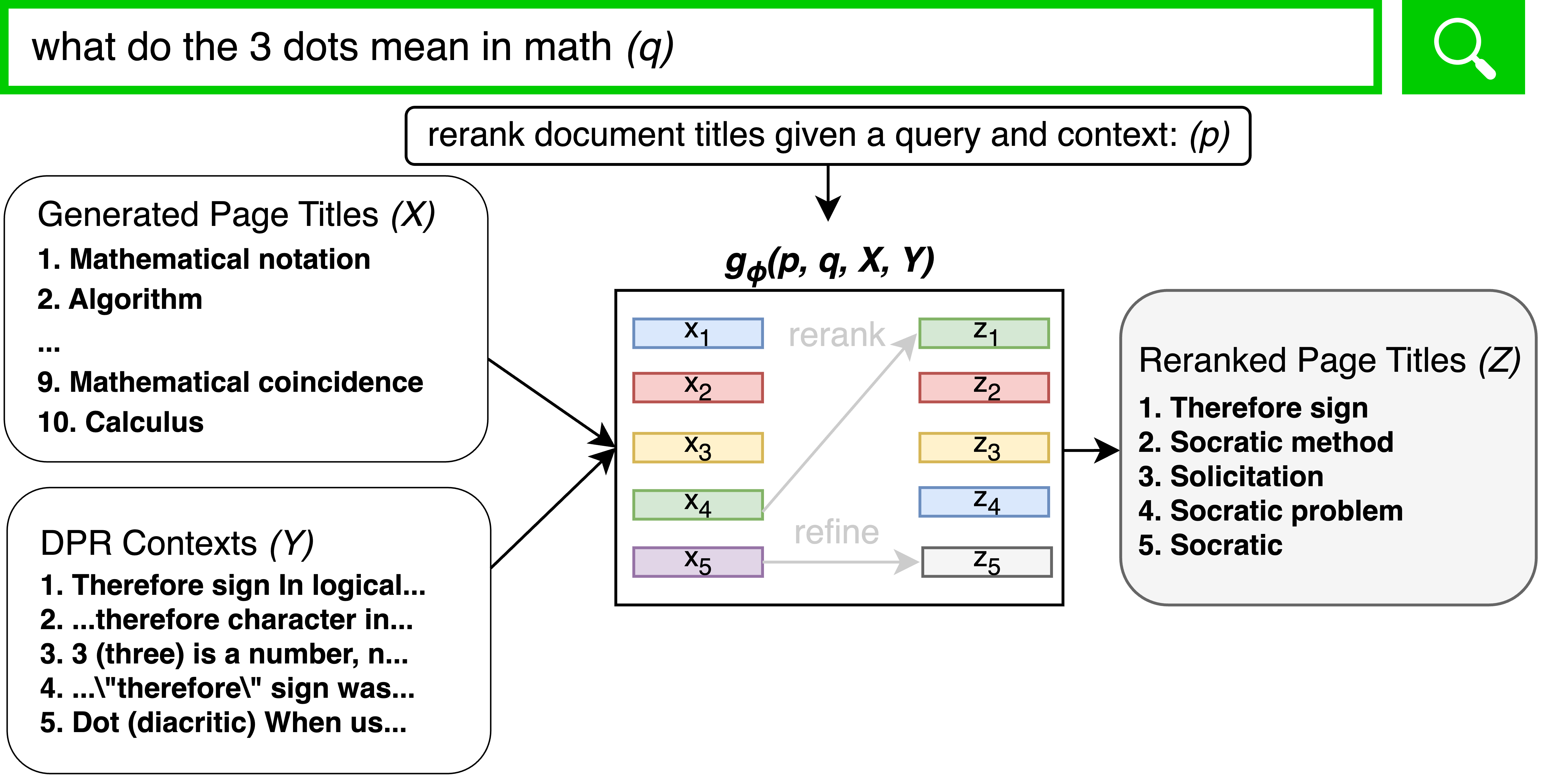}
    \caption{Re3val's Page Title Reranker ($g_\phi$) enhances generated page titles ($X$) with DPR contextual information ($Y$), producing reranked titles ($Z$). This is crucial when documents in $X$ lack a suitable answer to a query ($q$), as depicted in the figure.}
    \label{fig:abstract_img}
\end{figure}

Rather than using inner-product-based retrieval, generative retrieval models such as GENRE \citep{cao2021autoregressive} and CorpusBrain \citep{Chen_2022} generate page titles through constrained decoding, attaining higher R-Precision and Recall compared to DPR. In our work, we further evaluate how additional contextual information can benefit the generative retrieval models through reranking and how reinforcement learning can enhance relevance through reward signals. 

We introduce Re3val: Reinforced and Reranked Generative Retrieval, a novel framework specifically designed to address the challenges in neural information retrieval. Our approach utilizes 500k pre-training data and 48k task-specific data for training. Despite the reduced data used in distant supervision, Re3val achieves exceptional performance.
Our contributions are described as below:
\begin{itemize}
    \item{We minimize the entropy of the initially retrieved page titles with contexts obtained from DPR, facilitating the novel generative reranking process. Through this reranking procedure, Re3val outperforms other generative retrieval models, including GENRE, CorpusBrain, and SEAL \citep{bevilacqua2022autoregressive} in terms of average R-Precision across five tasks, showcasing an average increase of 1.9\%.}

    \item{We incorporate REINFORCE \citep{williams1992REINFORCE} to integrate information during the decoding process of generative retrieval. Combined with question generation, REINFORCE enables Re3val to outperform CorpusBrain zero-shot retrieval with an average improvement of 8\% in R-Precision across five tasks.}

    \item{We suggest a new generative "retrieve and read" pipeline that extracts the contexts for the reranked page titles, applies our context reranker, and grounds answers with the reranked contexts.  As a result, Re3val distinguishes itself by achieving the highest KILT scores among other generative retrieval models, with an average increase of 2.1\%.}
\end{itemize}

In summary, Re3val uses DPR contexts for reranking page titles, leading to improved R-Precision. Re3val enhances performance by integrating generated questions in pre-training and utilizing REINFORCE during distant supervision. Moreover, Re3val achieves more accurate answers by reading reranked contexts retrieved with the reranked page titles. These advancements enable Re3val to achieve state-of-the-art performance while also offering cost savings by reducing training time and minimizing the need for extensive data labeling.

\section{Related Work}

\subsection{Document Retrieval}
TF-IDF \citep{karen1972statistical} and BM25 \citep{robertson2009probabilistic} assign weight to terms in a document based on their term frequency and inverse document frequency. These methods cannot inherently consider semantic shift or distribution similarity while computing similarity metrics. In light of this limitation, \citet{karpukhin-etal-2020-dense} introduce the Dense Passage Retrieval (DPR), establishing a bi-encoder that creates dense embeddings of questions and related passages within a corpus. These embeddings are subsequently compared using a dot product operation. During inference, DPR retrieves the top-k relevant contexts employing either Nearest Neighbor Search or Maximum Inner Product Search on the FAISS index. \citet{guu2020retrieval} and \citet{lewis2020retrieval} retrieve knowledge from a corpus using DPR and generate an answer using a variant of the Transformer models. FiD (Fusion in Decoder) \citep{izacard-grave-2021-leveraging} extends T5 \citep{wolf-etal-2020-transformers} by combining independently encoded queries and retrieved passages to decode an answer. However, these models do not rerank retrieved documents that allow a reader to perform better with fewer contexts utilized for a reader.

\subsection{Generative Retrieval}
\citet{cao2021autoregressive} introduce an Autoregressive Entity Retrieval model (GENRE). GENRE utilizes seq2seq language models for page title retrieval and employs a trie-based constrained decoding approach. This allows GENRE to assign a probability of 0 to non-existing page titles, ensuring accurate retrieval. Moreover, \citet{Chen_2022} propose CorpusBrain, a generative retrieval model encoding the knowledge about the corpus through pre-training strategies. DEARDR \citep{thorne-2022-data} proposes three distinct pre-training regimens and a data-efficient distant supervision method for generative retrieval. Moreover, SEAL \citep{bevilacqua2022autoregressive} leverages an FM-Index to efficiently generate n-grams within the corpus for fast lookup speed without increasing the index size. The Differentiable Search Index (DSI) \citep{tay2022transformer} employs a seq2seq model to map individual queries to atomic document identifiers, which in turn are associated with segmented chunks of the document. Similarly, the Neural Corpus Index (NCI) \citep{wang2022neural} utilizes hierarchical k-means for document representation, generates queries based on content, and trains a transformer model with a Prefix-Aware Weight-Adaptive Decoder using Consistency-based regularization. However, these models overlook the opportunity to minimize additional entropies in retrieved page titles or documents by incorporating contextual information. Leveraging such information reduces randomness and refines the ranking. Moreover, these models overlook the potential benefits of harnessing knowledge during decoding. 

\subsection{Question Generation}
In the past, numerous endeavors \cite{labutov-etal-2015-deep, chali-hasan-2015-towards, serban-etal-2016-generating, duan-etal-2017-question} have been made to generate questions to enhance the task of Question Answering. Recently, studies analyzing questions have attempted to find the relationship with contexts. \citet{mao-etal-2021-generation} propose Generation-Augmented Retrieval (GAR) that generates query contexts. GAR employs a BM-25 retrieval model and achieves performance comparable to DPR. \citet{sachan-etal-2022-improving} create questions based on the retrieved contexts and rerank contexts based on the log-likelihood score over the generated questions. However, these studies overlook the fact that question generation can address the epistemic uncertainty arising from limited knowledge \citep{kendall2017what} in question answering tasks by minimizing the domain gap between pre-training and fine-tuning data.

\subsection{Reranking Models}
Reranking in information retrieval involves refining the initial ranking of retrieved documents by utilizing scores from a more complex query, as exemplified by Apache Solr\footnote{\url{https://solr.apache.org}}. Atlas \citep{izacard2022atlas} retrieves documents with Contriever \citep{izacard2022unsupervised}, reranks the retrieved documents, and reasons with FiD. Re\textsuperscript{2}G \citep{glass-etal-2022-re2g} employs a cross-encoder \citep{rosa2022defense, nogueira2020passage} to rerank retrieved documents based on softmax probability using $BM25(q) \cup DPR(q)$, determining the relevance between a query and context. FiD-Light \citep{Hofstatter2022fidlight} introduces a compression for encoded passages and reranks candidate lists using source pointers. These source pointers are textual indices that represent the relevant context, as initially introduced in FiD-Ex \citep{lakhotia-etal-2021-fid}. However, these reranking models do not perform reranking at the page title level and do not make use of a rerank query.

\subsection{Reinforcement Learning}
When framing text generation as a Reinforcement Learning (RL) problem, the state ($s_t$) represents the hidden states of the encoder and previously decoded outputs at time steps ${1, 2, ..., t - 1}$. The action ($a_t$) encompasses the encoding and decoding behaviors, as well as the decoded word at time step $t$ \citep{paulus2018a}. This formulation can incorporate non-differentiable feedback, such as common evaluation metrics as reward. Moreover, various RL methodologies such as REINFORCE \citep{williams1992REINFORCE}, Advantage Actor-Critic (A2C) \citep{mnih2016asynchronous}, and Proximal Policy Optimization (PPO) \citep{schulman2022proximal} are being successfully applied in a multitude of scenarios. This study primarily utilizes REINFORCE, a simple yet effective method.

\begin{figure*}[t]
    \centering
    \includegraphics[width=\textwidth]{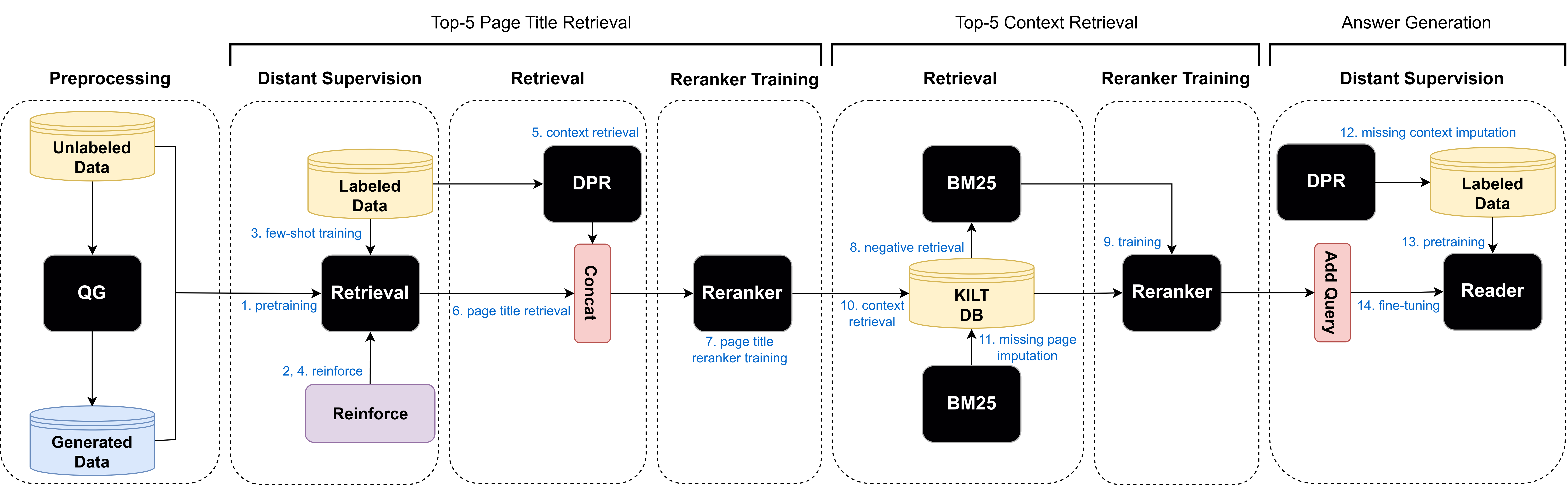}
    \caption{Re3val Training Pipeline. Generated questions after filtering are integrated into pre-training (1), followed by few-shot training (3) with REINFORCE (2, 4). Retrieved DPR contexts (5), perturbed page titles (6), and queries are concatenated for reranker training (7). Gold and negative passages retrieved with BM-25 are employed (8) for context reranker training (9). Contexts are retrieved using the top 5 reranked titles from KILT (10), where missing titles are imputed with BM-25 (11). DPR contexts are imputed (12) if lacking five gold contexts during FiD model pre-training (13). FiD model is fine-tuned using five reranked contexts (14).}
    \label{fig:training}
\end{figure*}
\begin{figure*}[t]
    \centering
    \includegraphics[width=\textwidth]{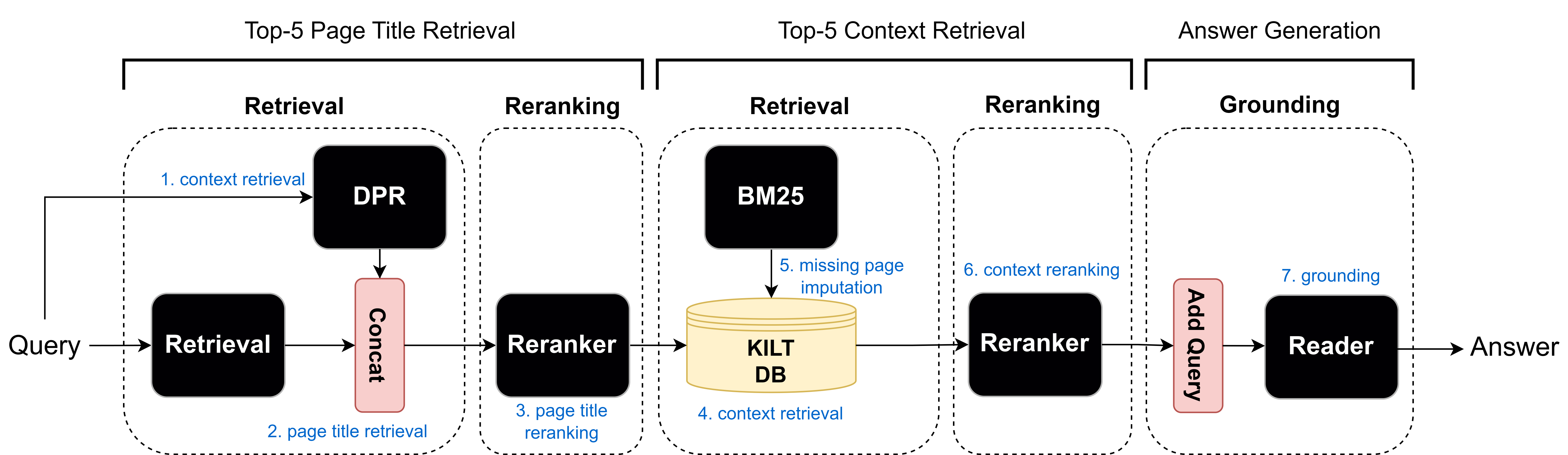}
    \caption{Re3val Inference Pipeline. Reranker concatenates retrieved DPR contexts (1), page titles (2), and query to rerank page titles (3). Contexts retrieved with the top five reranked page titles (4), including BM-25 imputed titles (5), are reranked (6). The top-5 reranked contexts are used to generate an answer (7).}
    \label{fig:inference}
\end{figure*}

\section{Methodology}
The primary contribution of Re3val is its capability to generatively rerank page titles by incorporating contextual information and to apply REINFORCE during distant supervision of a generative retrieval. Additionally, Re3val utilizes question generation for pre-training. Furthermore, Re3val pioneers the reading of contexts retrieved using page titles obtained through a generative retrieval approach.

The following elucidates the function of each component in Figure \ref{fig:training} with respect to its task.

\subsection{Page Title Retrieval (Stage 1-4)}
\paragraph{Distant Supervision (Stage 1,3)}
Following DearDr \citep{thorne-2022-data}, we pre-train the generative retrieval. To mitigate the domain shift problem during pre-training for question-answering and dialogue tasks, we generate questions for half of the pre-training passages. We utilize Flan-T5 base \citep{chung2022} to create questions given a prompt, "Generate a question related to the following Passage: ". Among generated questions, we employ Spacy's Entity Recognizer of en\_core\_web\_sm\footnote{\url{https://spacy.io}} to filter out ambiguous questions such as "Where is he". Specifically, we remove questions that do not contain entities other than DATE, MONEY, CARDINAL, TIME, QUANTITY, ORDINAL, and PERCENT. 

During the pre-training and fine-tuning of Re3val, an instructive prompt - "rank document titles given a query: " - is introduced before each query on the t5-small, t5-base, and t5-large \citep{wolf-etal-2020-transformers}. In Few-Shot training, we added labeled data to narrow the range of target candidates.

\paragraph{REINFORCE (Stage 2,4)}
A policy ($\pi$) is parameterized by $\theta$, where $T$ denotes the sequence length. Additionally, $R(\tau)$ signifies the cumulative reward associated with a trajectory $\tau$, characterized as a sequence of actions ($a$) and states ($s$). The formula for calculating the gradient of the REINFORCE objective function is:

\begin{equation}
\small
\nabla J(\theta) = E_{\pi_{\theta}}\left(\sum_{t=1}^{T}\nabla_{\theta}\log\pi_{\theta}(a_t, s_t)R(\tau)\right)
\end{equation}

The REINFORCE is employed during training to optimize the black box of zero-shot and few-shot retrieval in Re3val. The REINFORCE utilizes the R Precision of generated page titles as a reward. The effectiveness of the REINFORCE is demonstrated in Appendix \ref{sec:appendix:reinforce}
\subsection{Page Title Reranker (Stage 5-7)}
\label{app:rerankertrain}

Retrieved page titles are initially ranked based on their relevance score, computed by our retrieval model. Then, a reranking query can be introduced to refine the ranking further and increase the likelihood of obtaining the most relevant page titles. However, the KILT datasets do not provide a specific reranking query.

To address the limitation above, our page title reranker leverages contexts retrieved via an auxiliary index, such as the Dense Passage Retrieval multi-set checkpoint\footnote{\url{https://github.com/facebookresearch/DPR}}, to serve as the reranking query. 
Unlike the prompt for ranking, which is "rank document titles given a query: ", the prompt for reranking is modified to "rerank document titles given a query and contexts: ".

We have implemented a new training strategy to improve the refinement and reranking functions of our page title reranker. This strategy combines reinforced few-shot (Stage 4) and zero-shot (Stage 1) retrieved page titles during training. Additionally, we apply uniform shuffling to the page titles in the top half of the training sets generated by our zero-shot and few-shot retrieval.

Mixing titles from different checkpoints and shuffling retrieved page titles introduces noise to the input data. This noise is beneficial as it enables the page title reranker to filter out inconsistencies, outliers, and misleading patterns in the test set, ultimately enhancing its performance.

\subsection{Context Retrieval (Stage 10-11)}
\label{app:kiltdatabasetrain}

\paragraph{Preprocessing (Stage 10)}
To refine the data for context retrieval for a reader, we divide each context in the KILT Database into chunks, each consisting of 100 words. To ensure data quality and relevance, we filter out sentences that only contain a page title, as well as sentences containing the specific patterns, "Section::::" or "BULLET::::".

\paragraph{Extraction (Stage 10-11)}
After the page title reranking process, we acquire five reranked page titles. Subsequently, we retrieve the corresponding contexts for each page title. In situations where specific page titles are unavailable in the KILT database, we suggest using the BM-25 imputation method. This method employs the BM-25 algorithm to impute the most suitable page title from the KILT database. A detailed analysis of this imputation approach can be found in Appendix \ref{sec:appendix:imputation}.

\subsection{Context Reranker (Stage 8-11)}
To enhance the reader's experience, we reduce memory and context usage through our Context Reranker. Specifically, we use a cross-encoder to assess the relevance of a query and context pair for reranking the contexts derived from the five page titles. The input structure for our context reranker is as follows: "[CLS] Query [SEP] Context [SEP]".

We utilize gold passages as positive examples for training our Context Reranker on nboost/pt-bert-base-uncased-msmarco\footnote{\url{https://huggingface.co/nboost/pt-bert-base-uncased-msmarco}}. We also include two types of hard negative examples retrieved with BM-25: the top 128 unlabeled context chunks mapped to labeled page titles and the top 128 unlabeled context chunks mapped to the unlabeled page titles retrieved by our Page Title Reranker.

\subsection{Reader (Stage 12-14)}

We employ the Fusion in Decoder (FiD) as our reader for the reading task. During the pre-training phase of FiD, we utilize gold passages and impute DPR contexts for queries with fewer than five available gold contexts. Subsequently, following the pre-training phase, we perform fine-tuning of the FiD model using the top five or ten contexts retrieved by our context reranker.


\section{Experiments}
\subsection{Datasets}
We use datasets from the KILT \citep{petroni-etal-2021-kilt} benchmark. We study Natural Questions \citep{kwiatkowski-etal-2019-natural}, TriviaQA \citep{joshi-etal-2017-triviaqa}, and HotpotQA \citep{yang-etal-2018-hotpotqa} for question answering tasks, FEVER \citep{thorne-etal-2018-fever} for a fact-checking task, and WoW \citep{DBLP:journals/corr/abs-1811-01241} for a dialogue task, which are publicly available\footnote{\url{https://github.com/facebookresearch/KILT}}. Comprehensive details about the datasets are discussed in Appendix \ref{sec:appendix:data}.

\subsection{Evaluation}
\label{app:evaluation}
KILT utilizes a page-level retrieval strategy, and the assessment of page-level retrieval tasks measures the capacity to present a collection of Wikipedia pages as supporting evidence for a prediction, assessed through R-Precision and Recall@k metrics. R-Precision quantifies the proportion of relevant documents retrieved out of the total retrieved documents. However, Recall@k quantifies the proportion of relevant documents retrieved out of the total number of actual documents, taking into account only the top-k retrieved documents. Downstream reading tasks utilize different evaluation metrics depending on the specific task. For example, question-answering tasks are evaluated using Exact Match (EM) and F1 scores. Dialogue tasks employ metrics such as ROUGE-L and F1 scores. Fact verification tasks, on the other hand, are evaluated based on Accuracy. However, KILT has recently introduced the KILT score\footnote{\url{https://eval.ai/web/challenges/challenge-page/689/evaluation}} as a ranking metric for evaluating downstream performance. The KILT score takes into account post-processed Accuracy, EM, ROUGE-L, and F1 scores mentioned in Appendix \ref{sec:appendix:metrics:downstream}, but only if the R-Precision for a given query is 1. For detailed information regarding the metrics for evaluation, please refer to Appendix \ref{sec:appendix:metrics}.

\subsection{Page Title Retrieval}

\paragraph{Training}
We utilize 250k uniformly sampled June 2017 and August 2019 Wikipedia dumps for the pre-training phase across all datasets. Additionally, we generate questions from an additional 250k uniformly sampled Wikipedia dumps and include them in the training process. For fine-tuning, we utilize 48k uniformly sampled task-specific datasets. Detailed information about the datasets can be found in Appendix \ref{sec:appendix:data} and Table \ref{table:datainfo}. Importantly, we reinforce the zero and few-shot retrieval stages by employing the same dataset for each retrieval stage.

\paragraph{Evaluation}
We employ a multi-beam search approach with a beam size specified in Table \ref{tab:hyperparameter} to assess the performance on all development and test sets. In addition, we select the top five page titles from the list of multi-page titles generated per query for evaluation purposes.

\subsection{Page Title Reranker}
In our experimentation, we explore two types of initialization for our page title reranker. Firstly, we initialize the reranker using the plain t5-small, t5-base, and t5-large models. Secondly, considering the three different model sizes, we utilize the checkpoint from the reinforced few-shot retrieval process. To maintain input compatibility, we limit the query for the reranker's input to the first 250 words. In addition, the input - consisting of a query, ten page titles, and five contexts - is truncated to a maximum of 512 tokens.

\subsection{Context Reranker}
\label{app:contextreranker}
We input the first 150 words of a query for question-answering and fact-verification tasks. In the case of a dialogue task, the last 300 words of the query are used, as the final sentence often serves as the closure to the conversation. The maximum sequence length of input is detailed in Table \ref{tab:hyperparameter} and \ref{tab:configuration_wow}, providing further information on the specific limitations imposed on the input size.

\subsection{Reader}
Two types of inputs are used for pre-training our two versions of FiD. The first type includes only gold passages, while the second consists of gold passages and top-ranked Dense Passage Retrieval (DPR) contexts. For the Natural Questions (NQ) dataset, pre-training is conducted using the NQ FiD checkpoint, which has been pre-trained on 770 million parameters\footnote{\url{https://github.com/facebookresearch/FiD}}. For the remaining datasets, pre-training is performed using the TriviaQA FiD checkpoint, which has been pre-trained on 770 million parameters\footnotemark[7]. Regarding the WoW dataset, we retain the last 385 words of the query for input. For other datasets, we use the first 125 words. The maximum sequence length is outlined in Table \ref{tab:hyperparameter} and \ref{tab:configuration_wow}, providing specific details on the constraints imposed on input size.

An example of an input format is "question: query, title: page\_title, context: retrieved\_context". In this format, "question:", "title:", and "context:" are special tokens, while "query", "page\_title", and "retrieved\_context" represent variables denoting the respective components of the input.

\section{Result} 
\begin{table*}[h]
\centering
{\small

\begin{tabular}{ccccccccccc|cc}
\toprule
  & \multicolumn{6}{c}{Question Answering} & \multicolumn{2}{c}{Fact Check.} & \multicolumn{2}{c}{Dial.} & \multicolumn{2}{c}{Average}  \\
\textbf{Dataset} & \multicolumn{2}{c}{\textbf{NQ}} & \multicolumn{2}{c}{\textbf{TQA}} & \multicolumn{2}{c}{\textbf{HoPo}} & \multicolumn{2}{c}{\textbf{FEV}} & \multicolumn{2}{c}{\textbf{WoW}} & \\
\toprule
\textbf{Model} & \textbf{R-P} & \textbf{R@5} & \textbf{R-P} & \textbf{R@5} & \textbf{R-P} & \textbf{R@5} & \textbf{R-P} & \textbf{R@5} & \textbf{R-P} & \textbf{R@5} & \textbf{R-P} & \textbf{R@5} \\
\toprule
\multicolumn{13}{c}{\textbf{Zero-shot}}\\
\midrule
TF-IDF & 28.10 & - & 46.40 & - & 34.10 & - & 50.90  & - & 49.00 & - & 41.70 & -  \\
CorpusBrain & 28.25 & - & 42.76  & - & \textbf{44.84} & - &70.38 & - & 29.64 & - & 43.17 & - \\
\textbf{Re3val}$_{S}$ & 25.20 &29.62&\underline{47.24}&48.52&42.91&\underline{23.36}&74.99&84.19&52.31& 64.28 &48.53 & 49.99   \\
\textbf{Re3val}$_{B}$ & \underline{33.24} &\underline{37.90}&\textbf{47.25}&\underline{52.88}&\underline{43.82}&\textbf{24.79}&\underline{76.22}&\underline{83.42}&\textbf{56.45}&\underline{70.05}&\underline{51.40}&\underline{53.81}\\
\textbf{Re3val}$_{L}$ & \textbf{34.70} & \textbf{41.47} & 46.38 & \textbf{53.01} & 43.55 & 22.77 & \textbf{78.60} & \textbf{85.36} & \underline{55.67} & \textbf{72.77} & \textbf{51.78} & \textbf{55.07}   \\
\midrule
\multicolumn{13}{c}{\textbf{Few-shot (48k)}}\\

\midrule
\textbf{Re3val}$_{S}$ & 47.44& 49.20 &61.28&64.32&47.47&27.53&79.74&84.29&56.90 & 71.86& 58.57 & 59.44  \\
\textbf{Re3val}$_{B}$ &54.15&55.34&63.80&69.83&50.01&31.47& 78.67 & 82.47 & 62.00& 77.50 & 61.73 & 63.32  \\

\textbf{Re3val}$_{L}$ & 54.92 & 55.76 &  63.89 & 71.35 & 49.99 & 32.81 & 77.15 & 79.88 & 62.84 & 79.91 & 61.76 & 63.94   \\
\midrule
\multicolumn{13}{c}{\textbf{Full Fine-tuning}}\\
\midrule
DPR + BART & 54.29 & 65.52 & 44.49 & 56.99 & 25.04 & 10.40 & 55.33 & 74.29 & 25.48 & 55.10 & 40.93 & 52.46\\
RAG &59.49 & 67.06 & 48.68 & 57.13 & 30.59 & 12.59 & 61.94 & 75.55 & 57.78 & 74.63 & 51.70 & 57.39\\
GENRE & 60.25 & 61.36 & 69.16 & 75.07 & 51.27 & 34.03 & 83.64 & 88.15 & 62.88 & 77.74& 65.44 & 67.27 \\
KGI & 63.71 & \textbf{70.17} & 60.49 & 63.54 & - & - & 75.60 & 84.95 & 55.37 & 78.45 & - & -\\
SEAL & 63.16 & \underline{68.19} & 68.36 & \textbf{76.36} & \underline{58.83} & \textbf{51.03} & 81.45 & \underline{89.56} & 57.55 & 78.96 & 65.87 & \textbf{72.82} \\
TABi & 62.60 &  64.95 &\textbf{70.36} & 69.16 & 53.12 & 35.48 & \textbf{84.45} & 88.62 & 59.11 & 69.10 & 65.93 & 65.46\\
CorpusBrain & 60.32 & 61.21 & \underline{70.19} & \underline{75.64} & 51.80 & 34.57 & \underline{84.07} & \textbf{90.50} & \textbf{64.79} & \textbf{81.85} & 66.23 & 68.75 \\
\midrule
\multicolumn{13}{c}{\textbf{Reranking (48k)}}\\
\midrule
\textbf{Re3val}$_{S}$ & 59.63 & 60.78 & 59.84 & 64.43 & 54.93 & 38.50 & 81.22 & 85.90  & 56.90* & 71.86*  & 62.50 & 64.29 \\
\textbf{Re3val}$_B$ & \underline{64.75} & 63.05 & 66.31 & 71.95 & 56.65& 41.14 & 81.58 & 83.27 & 62.00* & 77.50* & \underline{66.26} & 67.38   \\
    \textbf{Re3val}$_L$ & \textbf{66.48} & 65.40 & 68.57 & 74.48 & \textbf{59.60} & \underline{44.21} & 82.78 & 85.71   & \underline{63.32} & \underline{79.88} & \textbf{68.15} & \underline{69.94}   \\
\bottomrule
\end{tabular}
}
\caption{The table above summarizes performance results for generative and bi-encoder retrieval models on KILT test sets. Top-performing models are highlighted in \textbf{bold}, and second-best in \underline{underline}. In Re3val, a reinforced version is used for Zero-shot and Few-shot (48k), while unreinforced version is used for Reranking (48k). Reranking (48k) involves a page title reranker trained using \textit{S} (t5-small), \textit{B} (t5-base), and \textit{L} (t5-large). For WoW dataset, reported scores are few-shot results, except {Re3val}$_L$, denoting the best overall result. Re$^2$G and FiD-Light are excluded as they perform reranking on a bi-encoder retrieval model using full data.
}
\label{tab:final_score}
\end{table*}

\subsection{Page Title Retrieval}
\label{app:result:retrieval}

\paragraph{Zero-shot Retrieval}

Based on the findings presented in Table \ref{tab:final_score}, CorpusBrain exhibits an 8\% lower R-Precision on average compared to Re3val, despite being trained on more than 500 times more data. We hypothesize that the question-generation process mitigates the epistemic uncertainty resulting from limited training data, thus minimizing the domain shift between the pre-training and task-specific fine-tuning data.

Examining Table \ref{tab:dev_final_score} in the Appendix, we observe that REINFORCE yields a modest improvement in the performance of zero-shot retrieval, with a few exceptions. Specifically, REINFORCE effectively captures the variability introduced during the constrained beam search exploration, as it utilizes the search results as a reward signal, thereby reducing bias towards the pre-training data in our retrieval model.

\paragraph{Few-shot Retrieval}
However, as indicated in Table \ref{tab:dev_final_score}, the effectiveness of REINFORCE diminishes when applied to the few-shot retrieval scenario. In some instances, REINFORCE results in performance degradation across specific datasets. We postulate that this phenomenon can be attributed to the inherent variance associated with Reinforcement Learning. Furthermore, the performance degradation may arise from the exploration-exploitation trade-off during the multi-beam search, where a broad range of solution spaces is explored, potentially leading to a decreased focus on exploitation. For instance, Appendix \ref{sec:appendix:recall} shows that the relative performance ranking can be reversed as the number of samples (K) increases.

\begin{table*}[h]
\centering
{\small

\begin{tabular}{ccccccccccc}
\toprule
&  & \multicolumn{6}{c}{Question Answering} & \multicolumn{1}{c}{Fact Check.} & \multicolumn{2}{c}{Dial.} \\
\textbf{Dataset} & \textbf{|C|} & \multicolumn{2}{c}{\textbf{NQ}} & \multicolumn{2}{c}{\textbf{TQA}} & \multicolumn{2}{c}{\textbf{HoPo}} & \multicolumn{1}{c}{\textbf{FEV}} & \multicolumn{2}{c}{\textbf{WoW}} \\
\toprule
\textbf{Model} & & K.-EM & K.-F1 & K.-EM & K.-F1 & K.-EM & K.-F1 & K.-AC & K.-RL & K.-F1  \\
\toprule
\multicolumn{11}{c}{\textbf{Pre-training (48k)}}\\
\midrule    
\textbf{Re3val} & 5 & 36.84 & 42.27 & 48.34 & 51.74 & 23.25 & 27.55 &  70.62 & 9.74 & 10.81 \\
\textbf{Re3val$_I$} & 5 & 39.88 & 45.43 & \underline{51.08} & 53.93 & 23.85 & 28.11 &  \textbf{73.09} & 9.88 & 11.08 \\
\midrule
\multicolumn{11}{c}{\textbf{Full Fine-tuning}}\\
\midrule
SEAL & 100 & 38.78 &	44.40 & 50.56 & \textbf{54.99} &	18.06 & 21.42 & 71.28 & 10.45 & 11.63	 \\
RAG & 5 & 32.69 & 37.91 & 38.13 & 40.15 & 	3.21 & 4.10 & 53.45	 & 7.59 & 8.75 \\
KGI & 5 & 36.36 & 41.83 & 42.85 & 46.08 & 	- & - & 64.41 & 10.36 & 11.79 \\
DPR + BART & 5 & 29.09 & 42.36 & 46.19 & 1.96 & 2.53 & 63.94 & 34.70 & 5.91 & 6.96 \\
\midrule
\multicolumn{11}{c}{\textbf{Few-shot (48k)}}\\
\midrule
\textbf{Re3val} & 5 & 38.92 & 45.06 & 50.05 & 53.14 & 23.94 & 28.26 & 71.06 & 11.70 & 13.46 \\
\textbf{Re3val} & 10 & 	\underline{40.17} & \textbf{46.53} & \textbf{51.31} & \underline{54.46} & 24.13 & 28.44 & 71.08 & 11.79 & 13.41 \\
\textbf{Re3val$_{I}$} & 5 & \textbf{40.44} & \underline{46.23} & 50.41 & 53.44 & \textbf{24.33} & \underline{28.64} & 72.78 & \textbf{12.01} & \underline{13.55}  \\
\textbf{Re3val$_{I}$} & 10 & 39.54 & 45.92 & 51.00 & 53.93 & \underline{24.22} & \textbf{28.71} & \underline{73.02} & \underline{11.94} & \textbf{13.57}  \\

\bottomrule
\end{tabular}
}
\caption{The final KILT scores of the test sets are reported above, as presented on the KILT Leaderboard. The best-performing models are indicated in \textbf{bold}, while the second-best models are \underline{underlined}. Additionally, the notation \textit{I} denotes the \textit{Imputation} of DPR contexts for missing gold contexts. |C| represents the number of contexts.
}
\label{tab:final_score_answer}
\end{table*}

\subsection{Page Title Reranker} 

The validity of our reranker's input concatenation is supported by the principles of Mutual Information theory \citep{learnedmiller2013entropy}. Let's define $X$ as the set of page titles and $Y$ as the set of DPR contexts, where $X$ takes values from $\mathcal{X} = \{x_1, x_2, ..., x_n\}$ and $Y$ takes values from $\mathcal{Y} = \{y_1, y_2, ..., y_n\}$. We denote the probability distribution of $X$ as $P(x)$.

The mutual information between $X$ and $Y$ is denoted as $I(X;Y)$, and it quantifies the amount of shared information between the two variables. It is calculated using the formula:

\begin{equation}
\small
I(X;Y) = \sum_{x \in \mathcal{X}} \sum_{y \in \mathcal{Y}} P(x, y) \log \frac{P(x, y)}{P(x)P(y)}
\end{equation}

By considering the joint probability of DPR contexts and page titles, $I(X;Y)$ allows us to gain insights into the dependency between these two variables. Therefore, our page title reranker leverages this shared information to reduce uncertainty in the ranking of page titles, thus improving the reranking and refinement process.

The results obtained from the dev sets are documented in Table \ref{tab:dev_final_score}. Table \ref{tab:dev_final_score} indicates that the page title reranker, fine-tuned from the reinforced few-shot retrieval, outperforms the reranker initialized from the T5 pre-trained model when the number of parameters is small. However, the opposite trend is observed as the number of parameters increases. While the knowledge about ranking compensates for the limited capacity to learn complex reranking patterns when the number of parameters is small, prior knowledge about ranking interferes with the reranking function as the number of parameters grows. In essence, ranking and reranking serve distinct purposes. Ranking focuses on sorting relevant documents, while reranking involves permuting the initially ranked documents.

The dialogue task requires more detailed reasoning over textual information than question-answering and fact-verification tasks. Reranking with a few parameters does not yield improvements in performance for the WoW test set, as indicated in Table \ref{tab:final_score}. Furthermore, the inconsistency between the test set results in Table \ref{tab:final_score} and the dev set results in Table \ref{tab:dev_final_score} for the reranking stage of the 770m, 770m parameter configuration highlights the need for further investigation.

\subsection{Context Reranker}
The performance of our Context Reranker, evaluated using gold passages and hard negative passages as described in Section \ref{app:contextreranker}, is presented in Table \ref{tab:context_dev}. Notably, our Context Reranker exhibits a higher precision compared to recall. This characteristic shows that the Context Reranker effectively filters out irrelevant and low-quality results, prioritizing accuracy in retrieving relevant documents, even if they may miss some. The high precision score indicates that relevant documents are ranked at the top. However, further investigation is required to examine the trade-off between precision and recall in the Context Reranker for downstream reading tasks.

\subsection{Reader}
The slight performance difference observed between the reader with 5 and 10 contexts in Table \ref{tab:final_score_answer} suggests that our context reranker excels in retrieving highly relevant documents at the top, showcasing its exceptional precision. Moreover, our context imputation pre-training strategy is effective, enabling Re3val to outperform SEAL, although SEAL utilizes 100 contexts for grounding with FiD. Finally, as indicated in Table \ref{tab:final_score_answer}, Re3val achieves superior results with only five passages, underscoring the advantages of our approach.

\section{Conclusion}
This paper presents Re3val, a novel reranking architecture for generative retrieval. Re3val achieves state-of-art performance with question generation, REINFORCE, and reranking. Succinctly, Re3val incorporates question generation to address epistemic uncertainty and domain shift. It utilizes REINFORCE on constrained beam search outputs to enhance exploration. Experimental results demonstrate Re3val's superiority over the CorpusBrain zero-shot baseline, with an average 8\% R-Precision improvement across five tasks using reduced pre-training data. Re3val also achieves an average 1.9\% R-Precision increase compared to other generative models via page title reranking with limited task-specific data. Moreover, by employing a context reranker before grounding, Re3val achieves top-1 KILT scores among generative retrieval models, showing an average 2.1\% improvement across five datasets. Re3val's data-efficient approaches reduce training time and labeling costs, representing notable advancements in generative retrieval.

\section*{Acknowledgement}
We express our gratitude to Professor Kee-Eung Kim and Huzama Ahmad from KAIST AI for providing valuable feedback and guidance during the implementation of REINFORCE. We appreciate ChatGPT 3.5's assistance in correcting writing errors. This work was supported by Institute of Information \& communications Technology Planning \& Evaluation (IITP) grant funded by the Korea government (MSIT)
(No.2019-0-00075, Artificial Intelligence Graduate School Program (KAIST)).

\section*{Limitations}
Given this project's time and resource limitations, a comprehensive comparison of REINFORCE with other reinforcement learning algorithms, such as PPO and TRPO, which require more memory for their reference model, is not feasible. Furthermore, the observed disparity between the performance on the development and test sets for both the retrieval and reader components necessitates further investigation. Lastly, it is worth noting that specific labeled page titles in the FEVER dataset are not present in the KILT database, introducing a discrepancy that should be considered.

\section*{Ethics Statement}
In this study, we utilize datasets obtained from various sources, including Natural Questions, TriviaQA, HotpotQA, FEVER, and Wizard of Wikipedia. These datasets serve as integral components of the KILT benchmark and are derived from the KILT knowledge source, which is based on the August 1st, 2019, Wikipedia dump. In addition to the 2019 Wikipedia dump, we incorporate the June 2017 Wikipedia dump into our pre-training. It is crucial to acknowledge that these datasets may contain instances of incorrect or misconstrued information, which could potentially result in the generation of biased, toxic, or fabricated content. Moreover, the utilization of language models, such as T5, during the training and preprocessing stages introduces the possibility of ethical risks that may be embedded within the internal parameters of these models. Consequently, it is imperative for researchers to exercise caution when employing our paper and the associated outputs and to establish suitable policies to mitigate any potential ethical risks that may arise from the use of these models in real-world production settings.

\bibliography{main}

\appendix

\section{Appendix}
\label{sec:appendix}

\subsection{Hyperparameters}
\label{sec:appendix:hyperparameters}
The default hyperparameter settings and hardware configurations employed for the overall tasks are outlined in Table \ref{tab:hyperparameter}, with further details provided in Tables \ref{tab:configuration_retrieval_reranker} to \ref{tab:configuration_epoch}. Given the limited hardware resources available in our academic environment, we utilize different GPUs for our models, as specified in Table \ref{tab:configuration_retrieval_reranker}. FiD, which uses ten passages, is trained with half of the batch size indicated in Table \ref{tab:hyperparameter} and \ref{tab:configuration_wow}.

\subsection{Data}
\label{sec:appendix:data}
The number of data points used for pre-training and fine-tuning the retrieval models for each task are outlined in Table \ref{table:datainfo}. GENRE and CorpusBrain utilize 21 billion data points from the 2019 Wikipedia dump and 9 billion from the Blink dataset. In the case of Re3val pre-training, we use a combination of the June 2017 and August 2019 Wikipedia dumps.

For tasks such as Natural Questions (NQ), Wizard of Wikipedia (WoW), TriviaQA, and FEVER, we pre-train the models using 125,000 samples from the 2017 Wikipedia dump and 125,000 relevant samples from the Wikipedia dump obtained through the Dense Passage Retrieval multi-set checkpoint. An additional 250,000 generated questions from the remaining samples are also included in NQ, WoW, and TriviaQA. For HotpotQA, we use 125,000 original contexts and 125,000 data points from the two Wikipedia dumps, generating questions with the remaining 125,000 original contexts and 125,000 data points from the Wikipedia dumps. All subsets are uniformly sampled.

For the Page Title reranking task, we utilize Hotpot contexts instead of Dense Passage Retrieval (DPR) contexts specifically for HotpotQA. For other tasks, we used the Dense Passage Retrieval multi-set checkpoint.

\subsection{Prefix Tree}
\label{sec:appendix:prefix-tree}

To construct and search the Prefix Tree for all tasks, we utilize the KILT knowledge source\footnote{\url{http://dl.fbaipublicfiles.com/KILT/kilt_knowledgesource.json}}. This knowledge source is employed as the basis for building and performing Trie Node search.

\subsection{Constrained Decoding}
\begin{figure}[t]
    \centering
    \includegraphics[width=0.8\linewidth]{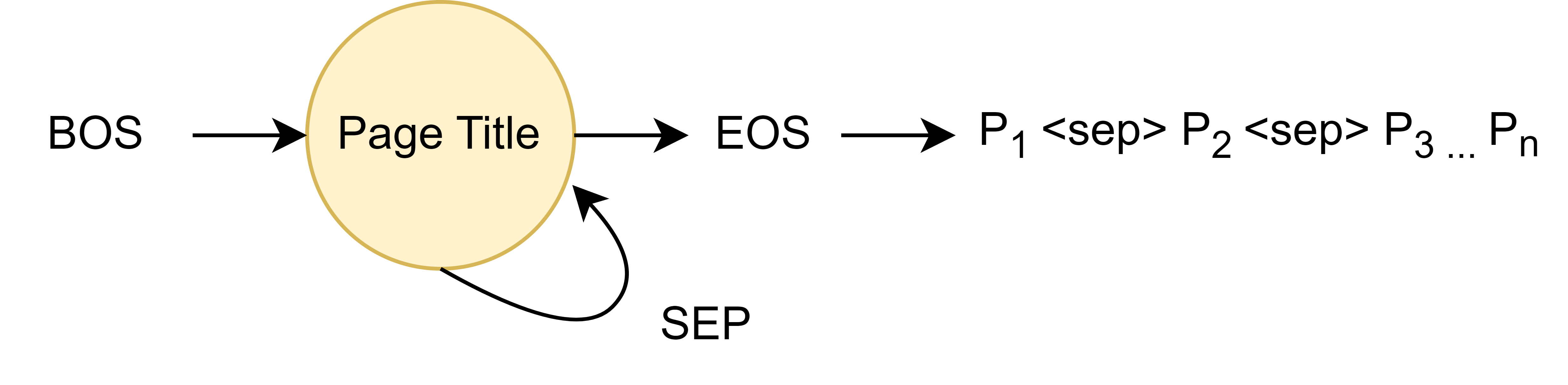}
    \caption{The decoding process in Re3val involves the utilization of DEARDR PTHL state machine decoding. During decoding, each page is conditionally decoded based on the previous page, as there are instances where multiple page titles are mapped to an answer. Furthermore, a query may have various answers, further influencing the decoding process.}
    \label{sec:appendix:fig:decode}
\end{figure}

In contrast to GENRE's constrained decoding \citep{cao2021autoregressive}, which predicts a single entity per beam, Re3val decodes a list of page titles per beam similar to DEARDR \citep{thorne-2022-data}, as depicted in Figure \ref{sec:appendix:fig:decode}. This approach enables us to capture the variability of related entities, as page titles are mapped to an answer in KILT datasets.

\subsection{REINFORCE}
\label{sec:appendix:reinforce}

This section presents a formal mathematical proof showcasing the optimization achieved by utilizing the REINFORCE algorithm in our retrieval system.

\subsubsection{Notation}

Let $J(\theta)$ denote the objective function. In the context of Re3val, $T$ represents the sequence length. The function $R(\tau)$ represents the return, which is the cumulative reward associated with a trajectory $\tau$, defined as a sequence of actions ($a$) and states ($s$). Finally, we denote the policy as $\pi$ with parameter $\theta$, and $\nabla$ represents the gradient operator.

\subsubsection{Proof}

The formula for computing the gradient of the REINFORCE objective function is given by:

\begin{equation}
\label{sec:appendix:reinforce:equation}
\nabla J(\theta) = E_{\pi_{\theta}}\left(\sum_{t=1}^{T}\nabla_{\theta}\log\pi_{\theta}(a_t, s_t)R(\tau)\right)
\end{equation}

The objective function (\ref{sec:appendix:reinforce:equation}) guides the policy $\pi_{\theta}$ towards the direction of the gradient. In equation (\ref{sec:appendix:reinforce:equation}), $R(\tau)$ is a scalar derived from the undifferentiable portion of Re3val, specifically the R-precision calculated using a constrained decoding prefix tree.

Re3val generates a sequence of page titles, represented as $\tau$, based on the policy $\pi$. The distribution of action $a$ given a state $s$ is denoted as $\pi_{\theta}(a|s)$. In the case of Re3val, a softmax function is applied to the cross entropy loss to obtain a probability distribution for the action $a$. Therefore, the policy parameter can be expressed as:

\begin{equation}
\log \pi_{\theta}(a_{t}, s_{t}) = \sum_{i=1}^{M} y_{i} \log \Bar{y_{i}}
\end{equation}

Here, $M$ represents the vocabulary size, which corresponds to the number of unique elements in the vocabulary.



In scenarios where $R(\tau_{1}) < R(\tau_{2})$, the model parameter undergoes a greater number of gradient updates in the direction of $\nabla_\theta (\sum_{j=1}^{M} \log\pi_{\theta}(a_t, s_t) R(\tau_{2}))$ compared to $\nabla_\theta (\sum_{j=1}^{M} \log\pi_{\theta}(a_t, s_t) R(\tau_{1}))$, provided that $R(\tau_{1}) > 0$ and $R(\tau_{2}) > 0$.

Consequently, the REINFORCE enhances the performance of zero-shot and few-shot retrieval by assigning more updates to samples that yield higher rewards, thereby promoting the learning of more relevant patterns and improving overall performance.

\subsection{Imputation}
\label{sec:appendix:imputation}

\subsubsection{Missing Page Imputation}
It has been observed that specific page titles retrieved by our model are absent in the KILT database despite applying the same preprocessing and tokenization procedures to these page titles as those utilized for building the Trie Node. This discrepancy in retrieval is systematically attributed to the labeler's mistake. Notably, as the missingness of top-ranked retrieved page titles can significantly impact performance, we assert that these page titles exhibit Missing Not At Random (MNAR) characteristics.

Let a dataset be $D = \{(x_t^{(i)}, o_t^{(i)})^{T_i}_{t=1}, y^{(i)}\}^n_{i=1}$ where $x$ be a page title, $o$ be a missing indicator, $y$ be a relevant context, $n$ be the number of data, $T$ be the number of page titles per a query, $f_\theta$ be Re3val's context reranker that produces a logit, and $k$ be the KILT database. For classification, $p(y|x_{1:T}, o_{1:T}, \theta) = \frac{e^{f_{\theta}(k(x_{1:T}, o_{1:T}))_1}}{\sum^{1}_{j=0}e^{f_{\theta}(k(x_{1:T}, o_{1:T}))_j}}$. Then, $p(x, o | \theta) = p(x|\theta)p(o|x,\phi)$, indicating missing ($o$) depends on both existing ($x$) and non-existing ($\phi$) page titles in the KILT database. That is, the probability of a missing retrieved page title in the database is related to the page title.

To address this MNAR missingness, we employ the BM-25 algorithm to impute the best matching page title from the KILT database. The outcomes of this imputation strategy are presented in Table \ref{tab:pt_title_imputation}, illustrating that the performance of our reranker on the test sets improves through the imputation.

\subsubsection{Missing Context Imputation}
Within the KILT dataset, contexts may be pertinent to an answer but have remained unlabeled due to biases from the labeler. This particular phenomenon aligns with the characteristics of Missing Not At Random (MNAR) since the absence of these contexts is systematically linked to the actions of the labeler. Table \ref{tab:final_score_answer} demonstrates a notable performance improvement when utilizing imputation techniques to address sparse contexts in a query using the DPR (Dense Passage Retrieval) method.

\subsection{KILT Leaderboard}
\label{sec:appendix:kilt_leaderboard}
Our performance results on the KILT downstream tasks can be found on the eval.ai leaderboard\footnote{\url{https://eval.ai/web/challenges/challenge-page/689/leaderboard}}. We prioritize the performance values reported in the original papers in Table \ref{tab:final_score} and \ref{tab:final_score_answer}. In cases where the original papers do not provide specific values, we rely on the results available on the KILT leaderboard. It is important to note that slight variations in the reported values may occur due to minor differences in the model versions used for evaluation across tasks.

\subsection{Metrics}
\label{sec:appendix:metrics}
\subsubsection{Page Title Retrieval}
Let us assume that $R$ represents the entire number of retrieved documents, and among these retrieved documents, $r$ is deemed relevant. In this case, R-Precision is the ratio of relevant retrieved documents to the entire number of retrieved documents, i.e., $\frac{r}{R}$. Similarly, Recall@k is calculated as $\frac{w}{n}$, the ratio of relevant retrieved documents to the entire number of actual documents, assuming there are $n$ actual documents and $w$ of these documents were successfully retrieved within a set of $k$ retrieved documents \cite{petroni-etal-2021-kilt}.

\subsubsection{Context Reranker}
Let us consider a classification task with the following definitions: TP (True Positive), TN (True Negative), FP (False Positive), and FN (False Negative). Precision is the ratio of true positives to the sum of true and false positives, given by $\frac{\text{TP}}{\text{TP + FP}}$. Similarly, Recall is defined as the ratio of true positives to the sum of true positives and false negatives, denoted as $\frac{\text{TP}}{\text{TP + FN}}$. The F1 score represents a balance between Precision and Recall, computed as the harmonic mean of the two metrics: $2 \times \frac{\text{Precision} \times \text{Recall}}{\text{Precision + Recall}}$. Accuracy, on the other hand, is calculated as the ratio of the sum of true negatives and true positives to the sum of true negatives, true positives, false positives, and false negatives, given by $\frac{\text{TP + TN}}{\text{TP + TN + FP + FN}}$.

\subsubsection{Reader}
\label{sec:appendix:metrics:downstream}
For the downstream reading task, we do not perform any post-processing on the gold and predicted outputs for the training and development sets. However, for the blind test sets, KILT applies post-processing techniques such as lowercase conversion, removal of articles, punctuation, and duplicate whitespace to the gold and predicted outputs. KILT maintains that these post-processing steps ensure consistency and fairness in the evaluation process.

\paragraph{KILT scores}
As mentioned in \ref{app:evaluation}, the KILT score incorporates post-processed Accuracy, EM, ROUGE-L, and F1 scores mentioned in Appendix \ref{sec:appendix:metrics:downstream}. However, these scores are considered only if the R-Precision for a given query is 1. The KILT scores provide a comprehensive evaluation of the system's performance on the KILT tasks by emphasizing high precision and relevance, in addition to other evaluation metrics.

\subsection{Recall Curve of the Page Title Reranker}
\label{sec:appendix:recall}
The plots below demonstrate the impact of different numbers of parameters on recall performance at varying levels of documents retrieved. A detailed discussion and analysis of these findings can be found in \ref{app:result:retrieval}
 of this paper.

\subsubsection{NQ}
\label{sec:appendix:recall:nq}
\begin{tikzpicture}[scale=.65]
\begin{axis}[
    xlabel={K},
    ylabel={Recall@K (\%)},
    xmin=1, xmax=10,
    ymin=55, ymax=85,
    xtick={1,2,3,4,5,6,7,8,9,10},
    ytick={50,55, 60,65, 70,75, 80, 85,90, 95,100},
  legend columns=-1,
  legend style={
    anchor=north,
    at={(0.5,-.2)}
    }         
  ],
  ymajorgrids=true,
    grid style=dashed,
]

\addplot[
    color=blue,
    ]
    coordinates {
    (1, 61.72)(2, 69.93)(3, 73.35)(4, 75.08)(5, 76)(6,76.52)(7, 76.84)(8,77.26)	(9, 77.69)(10, 77.93)
    };
\addplot[
    color=red,
    ]
    coordinates {
    (1, 63.66)(2, 71.73)(3, 74.87)(4, 76.28)(5, 77.44)(6,78.32)(7, 78.89)(8,79.49)	(9, 79.73)(10, 79.87)					
    };
\addplot[
    color=green,
    ]
    coordinates {
    (1, 65)(2, 72.82)(3, 75.47)(4, 76.95)(5, 78)(6,78.85)(7, 79.27)(8,79.91)	(9, 80.4)(10, 80.61)					
    };	
\addplot[
    color=pink,
    ]
    coordinates {
    (1, 56.36) (2, 67.43) (3, 71.2) (4, 73.53) (5, 74.52) (6, 75.4) (7, 75.85) (8, 76.31) (9, 76.67) (10, 76.98)				
    };	
\addplot[
    color=cyan,
    ]
    coordinates {
    (1, 66.3) (2, 73.6) (3, 76.42) (4, 78.5) (5, 79.1) (6, 79.77) (7, 80.37) (8, 80.54) (9, 80.93) (10, 81.14)				
    };
\addplot[
    color=magenta,
    ]
    coordinates {
    (1, 67.36) (2, 75.29) (3, 78.22) (4, 79.59) (5, 80.82) (6, 81.46) (7, 82.06) (8, 82.59) (9, 82.76) (10, 82.94)
    };	
\addlegendimage{blue, line legend} \addlegendentry{60m}
\addlegendimage{red, line legend} \addlegendentry{220m}
\addlegendimage{green, line legend} \addlegendentry{770m}
\addlegendimage{pink, line legend} \addlegendentry{2x60m}
\addlegendimage{cyan, line legend} \addlegendentry{2x220m} 
\addlegendimage{magenta, line legend}  \addlegendentry{2x770m} 

\end{axis}
\end{tikzpicture}
\subsubsection{TriviaQA}
\begin{tikzpicture}[scale=.65]
\begin{axis}[
    xlabel={K},
    ylabel={Recall@K (\%)},
    xmin=1, xmax=10,
    ymin=64, ymax=87,
    xtick={1,2,3,4,5,6,7,8,9,10},
    ytick={50,55, 60,65, 70,75, 80, 85,90, 95,100},
  legend columns=-1,
  legend style={
    anchor=north,
    at={(0.5,-.2)}
    }         
  ],
  ymajorgrids=true,
    grid style=dashed,
]

\addplot[
    color=blue,
    ]
    coordinates {
    (1, 64.75)(2, 75.91)(3, 79.6)(4, 80.72)(5, 81.64)(6,82.27)(7, 82.76)(8,83.15)	(9, 83.26)(10, 83.49)
    };
\addplot[
    color=red,
    ]
    coordinates {
    (1, 65.95)(2, 77.27)(3, 80.91)(4, 82.09)(5, 82.91)(6,83.3)(7, 83.75)(8,84.06)	(9, 84.36)(10, 84.7)					
    };									
\addplot[
    color=green,
    ]
    coordinates {
    (1, 66.77)(2, 77.07)(3, 80.84)(4, 82.03)(5, 82.98)(6,83.47)(7, 83.82)(8,84.01)	(9, 84.25)(10, 84.36)					
    };										
\addplot[
    color=pink,
    ]
    coordinates {
    (1, 65.25) (2, 75.01) (3, 78.19) (4, 78.99) (5, 80.07) (6, 80.63) (7, 81.0) (8, 81.28) (9, 81.51) (10, 81.84)				
    };	
\addplot[
    color=cyan,
    ]
    coordinates {
    (1, 66.95) (2, 77.44) (3, 80.91) (4, 81.96) (5, 83.04) (6, 83.6) (7, 84.06) (8, 84.31) (9, 84.61) (10, 84.79)				
    };
\addplot[
    color=magenta,
    ]
    coordinates {
    (1, 67.98) (2, 78.6) (3, 82.03) (4, 83.06) (5, 84.05) (6, 84.79) (7, 85.11) (8, 85.56) (9, 85.80) (10, 85.97)
    };	
\addlegendimage{blue, line legend} \addlegendentry{60m}
\addlegendimage{red, line legend} \addlegendentry{220m}
\addlegendimage{green, line legend} \addlegendentry{770m}
\addlegendimage{pink, line legend} \addlegendentry{2x60m}
\addlegendimage{cyan, line legend} \addlegendentry{2x220m} 
\addlegendimage{magenta, line legend}  \addlegendentry{2x770m} 

\end{axis}
\end{tikzpicture} \subsubsection{HotpotQA}
\begin{tikzpicture}[scale=.65]
\begin{axis}[
    xlabel={K},
    ylabel={Recall@K (\%)},
    xmin=1, xmax=10,
    ymin=35, ymax=67,
    xtick={1,2,3,4,5,6,7,8,9,10},
    ytick={30, 35, 40, 45, 50,55, 60,65, 70,75, 80, 85,90, 95,100},
  legend columns=-1,
  legend style={
    anchor=north,
    at={(0.5,-.2)}
    }         
  ],
  ymajorgrids=true,
    grid style=dashed,
]

\addplot[
    color=blue,
    ]
    coordinates {
    (1, 36.09)(2, 56.04)(3, 57.41)(4, 59.72)(5, 60.16)(6,61.26)(7, 61.53)(8,62.11)	(9, 62.22)(10, 62.62)
    };
    									
\addplot[
    color=red,
    ]
    coordinates {
    (1, 35.05)(2, 57.07)(3, 58.05)(4, 59.96)(5, 60.49)(6,61.38)(7, 61.63)(8,62.05)	(9, 62.15)(10, 62.6)					
    };	
    									
\addplot[
    color=green,
    ]
    coordinates {
    (1, 36.16)(2, 57.2)(3, 58.02)(4, 59.93)(5, 60.29)(6,61.3)(7, 61.49)(8,62.2)	(9, 62.34)(10, 62.8)					
    };																			
\addplot[
    color=pink,
    ]
    coordinates {
    (1, 36.18) (2, 56.01) (3, 57.16) (4, 59.37) (5, 59.91) (6, 60.86) (7, 61.13) (8, 61.74) (9, 61.88) (10, 62.26)
    };	
\addplot[
    color=cyan,
    ]
    coordinates {
    (1, 36.31) (2, 58.53) (3, 59.45) (4, 61.6) (5, 62.13) (6, 63.2) (7, 63.46) (8, 64.02) (9, 64.21) (10, 64.66)
    };
\addplot[
    color=magenta,
    ]
    coordinates {
    (1, 36.43) (2, 59.46) (3, 60.53) (4, 62.61) (5, 63.15) (6, 64.1) (7, 64.3) (8, 64.96) (9, 65.13) (10, 65.64)
    };
\addlegendimage{blue, line legend} \addlegendentry{60m}
\addlegendimage{red, line legend} \addlegendentry{220m}
\addlegendimage{green, line legend} \addlegendentry{770m}
\addlegendimage{pink, line legend} \addlegendentry{2x60m}
\addlegendimage{cyan, line legend} \addlegendentry{2x220m} 
\addlegendimage{magenta, line legend}  \addlegendentry{2x770m} 

\end{axis}
\end{tikzpicture} 
\subsubsection{FEVER}
\begin{tikzpicture}[scale=.65]
\begin{axis}[
    xlabel={K},
    ylabel={Recall@K (\%)},
    xmin=1, xmax=10,
    ymin=77, ymax=90,
    xtick={1,2,3,4,5,6,7,8,9,10},
    ytick={30, 35, 40, 45, 50,55, 60,65, 70,75, 80, 85,90, 95,100},
  legend columns=-1,
  legend style={
    anchor=north,
    at={(0.5,-.2)}
    }         
  ],
  ymajorgrids=true,
    grid style=dashed,
]

\addplot[
    color=blue,
    ]
    coordinates {
    (1, 81.84)(2, 85.75)(3, 87.7)(4, 88.34)(5, 88.86)(6,89.03)(7, 89.25)(8,89.36)	(9, 89.48)(10, 89.62)
    };
    										
\addplot[
    color=red,
    ]
    coordinates {
    (1, 77.87)(2, 79.6)(3, 81.09)(4, 81.39)(5, 81.77)(6,82.06)(7, 82.27)(8,82.45)	(9, 82.6)(10, 82.74)					
    };																		
\addplot[
    color=green,
    ]
    coordinates {
    (1, 79.15)(2, 81.44)(3,83.84)(4, 84.39)(5, 84.96)(6, 85.17)(7, 85.43)(8,85.5)	(9, 85.6)(10, 85.67)					
    };																												
\addplot[
    color=pink,
    ]
    coordinates {
    (1, 81.08) (2, 86.14) (3, 87.62) (4, 88.12) (5, 88.51) (6, 88.84) (7, 89.06) (8, 89.25) (9, 89.43) (10, 89.55)
    };
\addplot[
    color=cyan,
    ]
    coordinates {
    (1, 80.32) (2, 82.6) (3, 83.95) (4, 84.35) (5, 84.7) (6, 84.85) (7, 85.05) (8, 85.14) (9, 85.27) (10, 85.37)
    };
\addplot[
    color=magenta,
    ]
    coordinates {
    (1, 81.71) (2, 84.37) (3, 86.35) (4, 86.64) (5, 87.0) (6, 87.16) (7, 87.38) (8, 87.48) (9, 87.54) (10, 87.63)
    };
\addlegendimage{blue, line legend} \addlegendentry{60m}
\addlegendimage{red, line legend} \addlegendentry{220m}
\addlegendimage{green, line legend} \addlegendentry{770m}
\addlegendimage{pink, line legend} \addlegendentry{2x60m}
\addlegendimage{cyan, line legend} \addlegendentry{2x220m} 
\addlegendimage{magenta, line legend}  \addlegendentry{2x770m} 

\end{axis}
\end{tikzpicture}
\subsubsection{WoW}
\label{sec:appendix:recall:wow}
\begin{tikzpicture}[scale=.65]
\begin{axis}[
    xlabel={K},
    ylabel={Recall@K (\%)},
    xmin=1, xmax=10,
    ymin=40, ymax=76,
    xtick={1,2,3,4,5,6,7,8,9,10},
    ytick={30, 35, 40, 45, 50,55, 60,65, 70,75, 80, 85,90, 95,100},
  legend columns=-1,
  legend style={
    anchor=north,
    at={(0.5,-.2)}
    }         
  ],
  ymajorgrids=true,
    grid style=dashed,
]

\addplot[
    color=blue,
    ]
    coordinates {
    (1, 45.12)(2, 55.37)(3, 60.18)(4, 64.15)(5, 66.86)(6, 68.5)(7, 69.97)(8, 71.38)	(9, 72.27)(10, 73.41)
    };		
\addplot[
    color=red,
    ]
    coordinates {
    (1, 40.01)(2, 51.21)(3, 56.58)(4, 60.81)(5, 63.79)(6, 65.95)(7, 67.58)(8, 68.8)	(9, 69.45)(10, 70.2)					
    };
\addplot[
    color=green,
    ]
    coordinates {
    (1, 46.07)(2, 57.3)(3, 63.69)(4, 67.13)(5, 69.91)(6, 71.68)(7, 73.25)(8, 74.13)	(9, 74.66)(10, 75.25)	
    };																															
\addplot[
    color=pink,
    ]
    coordinates {
    (1, 42.53) (2, 51.31) (3, 56.02) (4, 59.14) (5, 61.53) (6, 63.23) (7, 65) (8, 66.27) (9, 67.29) (10, 68.5)
    };
\addplot[
    color=cyan,
    ]
    coordinates {
    (1, 47.18) (2, 54.03) (3, 58.28) (4, 60.84) (5, 63.23) (6, 65.52) (7, 66.9) (8, 68.21) (9, 69.12) (10, 69.68)
    };
\addplot[
    color=magenta,
    ]
    coordinates {
    (1, 46.07) (2, 57.3) (3, 62.84) (4, 66.34) (5, 69.25) (6, 70.92) (7, 72.82) (8, 73.9) (9, 74.79) (10, 75.57)
    };
\addlegendimage{blue, line legend} \addlegendentry{60m}
\addlegendimage{red, line legend} \addlegendentry{220m}
\addlegendimage{green, line legend} \addlegendentry{770m}
\addlegendimage{pink, line legend} \addlegendentry{2x60m}
\addlegendimage{cyan, line legend} \addlegendentry{2x220m} 
\addlegendimage{magenta, line legend}  \addlegendentry{2x770m} 

\end{axis}
\end{tikzpicture}

\begin{table*}[h]
\centering
{\small
\begin{tabular}{cccccccccccc}
\toprule \multicolumn{12}{c}{Question Answering}   \\
\multicolumn{4}{c}{\textbf{NQ}} & \multicolumn{4}{c}{\textbf{TQA}} & \multicolumn{4}{c}{\textbf{HoPo}} \\
\toprule
\textbf{PR} & \textbf{RC}  & \textbf{F1} & \textbf{AC} & \textbf{PR} & \textbf{RC} & \textbf{F1} & \textbf{AC}  & \textbf{PR} & \textbf{RC} & \textbf{F1} & \textbf{AC}  \\
\midrule
 62.04 & 21.10& 31.49 & 99.12 & 68.47 & 32.34 & 43.93 & 99.09 & 79.65 & 78.76 & 79.21 & 99.60 \\
\midrule
\end{tabular}
\begin{tabular}{cccccccc}
\toprule \multicolumn{4}{c}{Fact Check.} & \multicolumn{4}{c}{Dial.}   \\
\multicolumn{4}{c}{\textbf{FEV}} & \multicolumn{4}{c}{\textbf{WoW}} \\
\toprule
\textbf{PR} & \textbf{RC}  & \textbf{F1} & \textbf{AC} & \textbf{PR} & \textbf{RC} & \textbf{F1} & \textbf{AC} \\
\midrule
76.56 & 54.35 & 63.57 & 99.59 & 63.45 & 7.69 & 13.72 & 99.56 \\
\bottomrule
\end{tabular}
}
\caption{The results of our Context Reranker on the dev sets are presented in terms of Precision (PR), Recall (RC), Accuracy (AC), and F1-Score (F1).}
\label{tab:context_dev}
\end{table*}

\begin{table*}[h]
\centering
{\small
\begin{tabular}{cccccccccc}
\toprule
\textbf{Configuration} & \textbf{Retrieval$_L$} & \textbf{Reranker$_L$} &  \textbf{Reranker2} & \textbf{FiD}  \\
\midrule
learning rate &  5e-4 &  5e-4 &5e-5 & 1e-4 \\
scheduler &  constant w/ warmup &  constant w/ warmup & linear & constant \\
warmup ratio &  10\% &  10\% & 0 & 0 \\
eval steps ratio &  10\% &  10\% & 10\% & 10\% \\
batch size & 46* & 10  &  1200* & 32*  \\
max seq length &  200* &  512 & 250* & 250* \\
max target length &  30 &  30 & 50 & 50 \\
epoch & 5*  & 10* & 4 & 5*  \\
train beam size & 1  & 1 & 1 & 1  \\
eval beam size & 10  & 10 & 1 & 1  \\
test beam size & 5  & 5 & 1 & 1  \\
dropout rate & 0.2 & 0.2  &  0 & 0  \\
optimizer & AdamW & AdamW& AdamW& AdamW \\
gpu & RTX6000 & RTX6000 & A100 & A100 \\
early stopping steps & 4  & 4 & 4 & 4  \\

\bottomrule
\end{tabular}
}
\caption{The hyperparameter and hardware configurations used in our study are described above. The "Reranker" refers to the page title reranker, while "Reranker2" represents the context reranker. The asterisks (*) denote cases where different values were used for specific tasks. Further information can be found in Tables \ref{tab:configuration_retrieval_reranker} to \ref{tab:configuration_epoch}.}
\label{tab:hyperparameter}
\end{table*}

\begin{table*}[h]
\centering
{\small
\begin{tabular}{cccccccccccc}
\toprule
\textbf{Configuration} & \textbf{Retrieval$_S$}&  \textbf{Retrieval$_B$} & \textbf{Retrieval$_L$} & \textbf{Reranker$_S$} & \textbf{Reranker$_B$} & \textbf{Reranker$_L$} \\
\midrule

batch size & 220 & 160 & 46 & 70 & 35 &10    \\
gpu & RTX4000 & RTX3090 & RTX6000 & RTX4000 & RTX6000 & RTX6000 \\

\bottomrule
\end{tabular}
}
\caption{The retrieval and reranker models were configured differently with varying numbers of parameters.}
\label{tab:configuration_retrieval_reranker}
\end{table*}

\begin{table*}[h]
\centering
{\small
\begin{tabular}{ccccccccccc}
\toprule
\textbf{Configuration} & \textbf{Retrieval$_S$}&  \textbf{Retrieval$_B$} & \textbf{Retrieval$_L$}  & \textbf{Reranker2} & \textbf{FiD} \\
\textbf{Dataset} & WoW & WoW & WoW & WoW & WoW\\
\midrule

batch size & 110 & 95 & 20 & 600 & 16    \\
max seq length & 512 & 512 & 512 & 500 & 500     \\

\bottomrule
\end{tabular}
}
\caption{The configuration for the Wizard of Wikipedia (WoW) dataset is adjusted to accommodate the longer length of the input.}
\label{tab:configuration_wow}
\end{table*}

\begin{table*}[h]
\centering
{\small
\begin{tabular}{cccccccccccccc}
\toprule
\textbf{Configuration} & \multicolumn{2}{c}{\textbf{Retrieval}}&   \multicolumn{3}{c}{\textbf{Reranker}} & \textbf{FiD} \\

\textbf{Dataset} & FEV & WoW & NQ & FEV & WoW & TQA \\

\midrule

epoch & 1 & 1 & 20 & 1 & 1 & 1  \\
\bottomrule
\end{tabular}
}
\caption{Different configurations are utilized for certain datasets, deviating from the settings outlined in \ref{tab:hyperparameter}.}
\label{tab:configuration_epoch}
\end{table*}

\begin{table*}[h]
\centering
{\small
\begin{tabular}{cccccc}
\toprule
\textbf{Model} & \textbf{NQ} & \textbf{TQA} & \textbf{HoPo} & \textbf{FEV} & \textbf{WoW} \\
\midrule
\multicolumn{6}{c}{\textbf{Pre-training}}\\
\midrule
\textbf{Re3val} & 500,000 & 500,000 & 500,000 & 250,000 & 500,000 \\
GENRE & 30,000,000 & 30,000,000 & 30,000,000 & 30,000,000 & 30,000,000 \\
CorpusBrain & 30,000,000 & 30,000,000 & 30,000,000 & 30,000,000 & 30,000,000 \\
\midrule
\multicolumn{6}{c}{\textbf{Fine-tuning}}\\
\midrule
\textbf{Re3val} & 48,000 & 48,000 & 48,000 & 48,000 & 48,000  \\
GENRE & 87,372 & 61,844 & 88,869 & 104,966 & 63,734  \\
CorpusBrain & 87,372 & 61,844 & 88,869 & 104,966 & 63,734 \\
\bottomrule
\end{tabular}
}
\caption{The number of datasets utilized for training in our approach is smaller than that employed by other generative retrieval models.}
\label{table:datainfo}
\end{table*}

\begin{table*}[h]
\centering
{\small

\begin{tabular}{ccccccccccc|cc}
\toprule
  & \multicolumn{6}{c}{Question Answering} & \multicolumn{2}{c}{Fact Check.} & \multicolumn{2}{c}{Dial.} & \multicolumn{2}{c}{Average}  \\
\textbf{Dataset} & \multicolumn{2}{c}{\textbf{NQ}} & \multicolumn{2}{c}{\textbf{TQA}} & \multicolumn{2}{c}{\textbf{HoPo}} & \multicolumn{2}{c}{\textbf{FEV}} & \multicolumn{2}{c}{\textbf{WoW}} & \\
\toprule
\textbf{Model} & \textbf{R-P} & \textbf{R@5} & \textbf{R-P} & \textbf{R@5} & \textbf{R-P} & \textbf{R@5} & \textbf{R-P} & \textbf{R@5} & \textbf{R-P} & \textbf{R@5} & \textbf{R-P} & \textbf{R@5} \\
\toprule
\multicolumn{13}{c}{\textbf{Before Imputation}}\\
\midrule
\textbf{Re3val}$_{S}$ & 59.00 & \textbf{61.97} & 59.69 & 64.29 & 54.70 & 38.18 & 81.22 & 85.90  &  56.90* & 71.86*& 62.30 & \textbf{64.44} \\

\textbf{Re3val}$_B$ & 64.75 & 63.05 & 66.29 & 71.93 & 55.76& 39.59 & 81.58 & 83.27 & 62.00* & 77.50* & 66.01 & 66.67   \\

\textbf{Re3val}$_L$ & 66.48 & 65.40 & 68.55 & 74.47 & 59.58 & 44.21 & 82.29 & 85.25   & 63.32 & 79.88 & 67.94 & 69.13   \\

\midrule
\multicolumn{13}{c}{\textbf{After Imputation}}\\
\midrule
\textbf{Re3val}$_{S}$ & \textbf{59.63} & 60.78 & \textbf{59.84} & \textbf{64.43} & \textbf{54.93} & \textbf{38.50} & 81.22 & 85.90  & 56.90* & 71.86*  & \textbf{62.50} & 64.29 \\

\textbf{Re3val}$_B$ & 64.75 & 63.05 & \textbf{66.31} & \textbf{71.95} & \textbf{56.65}& \textbf{41.14} & 81.58 & 83.27 & 62.00* & 77.50* & \textbf{66.26} & \textbf{67.38}   \\
    \textbf{Re3val}$_L$ & 66.48 & 65.40 & 68.55 & 74.47 & \textbf{59.60} & 44.21 & \textbf{82.37} & 85.25   & 63.32 & 79.88 & \textbf{68.06} & 69.13   \\

\bottomrule
\end{tabular}
}
\caption{The impact of page title imputation using BM-25.}
\label{tab:pt_title_imputation}
\end{table*}

\begin{table*}[h]
\centering
{\small

\begin{tabular}{ccccccccccc}
\toprule
&  & \multicolumn{6}{c}{Question Answering} & \multicolumn{1}{c}{Fact Check.} & \multicolumn{2}{c}{Dial.} \\
\textbf{Dataset} & \textbf{|P|} & \multicolumn{2}{c}{\textbf{NQ}} & \multicolumn{2}{c}{\textbf{TQA}} & \multicolumn{2}{c}{\textbf{HoPo}} & \multicolumn{1}{c}{\textbf{FEV}} & \multicolumn{2}{c}{\textbf{WoW}} \\
\toprule
\textbf{Model} & & EM & F1 & EM & F1 & EM & F1 & AC & RL & F1  \\
\toprule
\multicolumn{11}{c}{\textbf{Few-shot (48k)}}\\
\midrule
\textbf{Re3val} & 5 & 39.06 & 48.58 & 40.49 & 50.54 & 35.13 & 45.60 & 88.25 & 17.06 & 17.49 \\
\textbf{Re3val$_I$} & 5 & \textbf{41.50} & 51.02 & 40.98 & 51.15 &  \underline{36.27} & \textbf{47.15} & \underline{89.83} & \underline{17.68} & \underline{17.87} \\
\textbf{Re3val} & 10 & 40.36 & \underline{51.15} & \underline{42.84} & \underline{53.29} & 35.09 & 46.02 & 88.42 & 17.22 & 17.56 \\
\textbf{Re3val$_I$} & 10 & \underline{41.35} & \textbf{51.84} & \textbf{43.35} & \textbf{53.74} & \textbf{36.30} & \underline{46.93} & \textbf{90.09} & \textbf{17.83} & \textbf{17.90} \\
\bottomrule
\end{tabular}
}
\caption{The best scores achieved on the dev sets when fine-tuning FiD are presented in the table above. The values highlighted in \textbf{bold} indicate the best scores, while those \underline{underlined} indicate the second-best scores. The notation \textit{I} represents the \textit{Imputation} of DPR contexts for missing gold contexts. }
\end{table*}

\begin{table*}[h]
\centering
{\small

\begin{tabular}{ccccccccccc}
\toprule
&  & \multicolumn{6}{c}{Question Answering} & \multicolumn{1}{c}{Fact Check.} & \multicolumn{2}{c}{Dial.} \\
\textbf{Dataset} & \textbf{|P|} & \multicolumn{2}{c}{\textbf{NQ}} & \multicolumn{2}{c}{\textbf{TQA}} & \multicolumn{2}{c}{\textbf{HoPo}} & \multicolumn{1}{c}{\textbf{FEV}} & \multicolumn{2}{c}{\textbf{WoW}} \\
\toprule
\textbf{Model} & & EM & F1 & EM & F1 & EM & F1 & AC & RL & F1  \\
\toprule
\multicolumn{11}{c}{\textbf{Pre-training (48k)}}\\
\midrule
\textbf{Re3val} & 5 & 44.88 & 52.86 & 62.24 & 67.17 & 31.78 & 40.78 & 86.30 & 14.53 & 15.89 \\
\textbf{Re3val$_I$} & 5 & 48.75 & 56.58 &	66.23 &	70.65 & 33.90 & 43.49 & 	89.43 & 14.74 & 16.36 \\
\midrule
\multicolumn{11}{c}{\textbf{Full Fine-tuning}}\\
\midrule
SEAL & 100 & 		\textbf{53.74} &	\textbf{62.24} & 	\underline{70.86}	&  \textbf{77.29} &		\textbf{40.46}	& \textbf{51.44} & \underline{89.54} & 	16.65	& 18.34	 \\
RAG & 5 & 	44.39 &	52.35 & 	\textbf{71.27} &	\underline{75.88} & 	26.97 &	36.03 & 86.31	 & 	11.57	& 13.11 \\
KGI & 5 & 	45.22 &	53.38 & 	60.99 &	66.55 & 	- & - & 	85.58 & 	16.36	& 18.57 \\
DPR + BART & 5 & 	39.75 & 		48.43 &	59.60	& 66.53 &	31.77 &	41.56 & 86.32 & 	13.27	& 15.12 \\
\midrule
\multicolumn{11}{c}{\textbf{Few-shot (48k)}}\\
\midrule
\textbf{Re3val} & 5 & 47.92 & 56.46 & 64.39	& 69.14 &35.39 & 45.04 & 	87.36 & 16.75 & 19.03 \\
\textbf{Re3val} & 10 & 	\underline{49.79}	& \underline{58.94} & 66.57 & 71.42 &35.73	&45.48 & 	87.15 & 16.92 &	18.93 \\
\textbf{Re3val$_I$} & 5 & 49.58 & 57.75 & 65.06 & 69.96 & 36.45 & 46.66 & 89.27 & \textbf{17.10} & \underline{19.06} \\
\textbf{Re3val$_I$} & 10 & 	48.68 &	57.37 & 65.87 &	70.49 & 	\underline{36.52}	& \underline{46.89} & \textbf{89.59} & \underline{17.06} &	\textbf{19.16} \\
\bottomrule
\end{tabular}
}
\caption{Reader scores of test sets on the KILT Leaderboard. The \textbf{bolded} are the best and the \underline{underlined} are the second best. \textit{I} indicates the \textit{Imputation} of DPR contexts for missing gold contexts. Note that the reader scores are not final scores as final scores are the KILT scores which award reader scores if R-Precision is 1.
}
\label{tab:final_reader_test}
\end{table*}

\begin{table*}[h]
\centering
{\small
\begin{tabular}{ccccccccccccc}
\toprule & &
  & \multicolumn{6}{c}{Question Answering} & \multicolumn{2}{c}{Fact Check.} & \multicolumn{2}{c}{Dial.}   \\
\textbf{Dataset} & & &  \multicolumn{2}{c}{\textbf{NQ}} & \multicolumn{2}{c}{\textbf{TQA}} & \multicolumn{2}{c}{\textbf{HoPo}} & \multicolumn{2}{c}{\textbf{FEV}} & \multicolumn{2}{c}{\textbf{WoW}} \\
\toprule
\textbf{Model} &\textbf{|P|} &\textbf{Stage} & \textbf{R-P} & \textbf{R@5} & \textbf{R-P} & \textbf{R@5} & \textbf{R-P} & \textbf{R@5} & \textbf{R-P} & \textbf{R@5} & \textbf{R-P} & \textbf{R@5}  \\
\midrule
\textbf{Re3val} & 60m & \textit{Z} & $26.40$ & $35.35$ & 45.62 & 59.38 & 52.95 & 45.91 & 77.70 & 84.93 &\underline{46.40} & 58.91   \\
\textbf{Re3val} & 60m & \textit{Z}, \textit{P} & $27.42$ & $36.02$ & 46.05 & 58.95 & 52.67 & 45.94 & 78.49 & 85.92 &44.27 & 56.81  \\
\textbf{Re3val} & 60m & \textit{F} & $45.40$ & $60.49$ & 59.49 & 71.99 &51.06 & 49.45 & 81.74 & 87.73 & \textbf{48.10} & \textbf{67.62}  \\
\textbf{Re3val} & 60m & \textit{F}, \textit{P} & $47.59$ & $62.18$ & 60.68 & 73.00 & 50.45 & 49.59 & 81.90 & 87.60 & 46.23 & 65.88  \\
\textbf{Re3val} & 60m & \textit{R} & \underline{61.72} & \underline{76.00} & \underline{64.75} & \textbf{81.64} & 56.79 & \underline{60.16} & \textbf{84.79} & \textbf{88.86} & 45.12 & 66.86 \\ 
\textbf{Re3val} & 60m & \textit{R}, \textit{P} & \textbf{62.39} & $75.36$ & 63.78 & \underline{81.36} & \textbf{57.39} & \textbf{60.32} & \textbf{84.79} & 88.07 &43.98 & \underline{67.13} \\
\midrule
\textbf{Re3val} & 60m,60m & \textit{R} & 56.36 & 74.52 & \textbf{65.25} & 80.07 & \underline{57.04} & 59.91 & \underline{83.87} & \underline{88.51} & 42.53 & 61.53 \\
\textbf{Re3val} & 60m,60m & \textit{R}, \textit{P} &  61.37 & \textbf{76.67}  &  64.43 & 80.29  &  56.72 & 59.73  & 82.94 & 87.93  &  36.97 & 58.32  \\
\midrule
\textbf{Re3val} & 220m & \textit{Z} & 32.78 & 45.93 & 47.02 & 62.72 & 52.29 & 46.78 & 72.27 & 85.98  & 49.84 & 60.31  \\
\textbf{Re3val} & 220m & \textit{Z}, \textit{P} & $35.78$ & $47.97$ & 42.40 & 60.59 & 54.13 & 47.64 & 77.25 & 86.81 & 49.18 & 61.85  \\
\textbf{Re3val} & 220m & \textit{F} & 54.74 & $69.05$ &61.90 & 77.87 & 50.69 & 51.97 & 79.15 & 82.58 & \underline{52.00} & \underline{71.77}   \\
\textbf{Re3val} & 220m & \textit{F}, \textit{P} & 54.35 & $68.56$ & 61.78 & 78.52 & 50.43 & 51.88 & 78.74 & 81.95 & \textbf{52.72} & \textbf{72.10} \\ 
\textbf{Re3val} & 220m & \textit{R} & $63.66$ & $77.44$ & \underline{65.95} & \underline{82.91} & 57.54 & 60.49 & 79.82 & 81.77  & 40.01 & 63.79 \\ 
\textbf{Re3val} & 220m & \textit{R}, \textit{P} & 64.22 & 76.35 & 65.80 & 82.87 & 57.69 & 60.39 & 79.86 & 82.52 & 39.06 & 62.41 \\ 
\midrule
\textbf{Re3val} & 220m,220m & \textit{R} & \textbf{66.30} & \textbf{79.10} & \textbf{66.95} & \textbf{83.04} & \textbf{58.85} & \textbf{62.13} & \underline{82.39} & \textbf{84.70} & 47.18 & 63.23 \\
\textbf{Re3val} & 220m,220m & \textit{R}, \textit{P} &  \underline{65.67} & \underline{78.43}   &  64.51 & 80.71  &  \underline{58.73} & \underline{61.82}  &  \textbf{82.84} & \underline{84.59}  &  39.06 & 62.38  \\
\midrule
\textbf{Re3val} & 770m & \textit{Z} & $32.11$ & $47.83$ & 43.37 & 61.19 & 48.10 & 46.33 & 78.73 & 83.77 & 49.67 & 65.55  \\
\textbf{Re3val} & 770m & \textit{Z}, \textit{P} & $33.84$ & $49.77$ & 44.95 & 63.22 & 46.24 & 44.90 & 81.08 & \textbf{87.94} & 50.36 & 65.19   \\
\textbf{Re3val} & 770m & \textit{F} & $55.97$ & $71.24$ & 64.06 & 79.92 & 50.39 & 51.85 & 80.46 &  82.97 & \textbf{55.34} & \textbf{74.89}   \\
\textbf{Re3val} & 770m & \textit{F}, \textit{P} & $57.00$ & $71.23$ & 63.61 & 79.79 & 50.62 & 52.27 & 79.40 & 82.40 & \underline{53.90} & \underline{74.36}  \\
\textbf{Re3val} & 770m & \textit{R} & \underline{65.00} & $78.00$ & 66.77 & \underline{82.98} & 57.66 & 60.29 & \underline{81.64} & 84.96 & 46.07 & 69.91  \\
\textbf{Re3val} & 770m & \textit{R}, \textit{P} & $64.65$ & \underline{78.22} & \underline{67.25} & 81.82 & 57.95 & 60.48 & 81.26 & 84.74 & 38.47 & 62.38  \\
\midrule
\textbf{Re3val} & 770m,770m & \textit{R} & \textbf{67.36} & \textbf{80.82} & \textbf{67.98} & \textbf{84.05} & \underline{59.75} & \underline{63.15} & \textbf{84.68} & \underline{87.00} & 46.07 & 69.25 \\
\textbf{Re3val} & 770m,770m & \textit{R}, \textit{P} &  63.80 & 77.79  &  65.05 & 79.79  & \textbf{59.76} & \textbf{63.26}  &  81.43 & 82.77  & 46.73 & 69.68 \\
\bottomrule
\end{tabular}
}
\caption{The performance of the development sets is evaluated at each stage of the training, considering different numbers of parameters. The stages include zero-shot retrieval (Z), few-shot retrieval (F), reranking (R), and reinforcement (P). The parameter counts |P| represent the parameters used to train the retrieval and reranker models. The comma (,) in |P| indicates that the retrieval and reranker were initialized separately. In contrast, the absence of a comma (,) signifies that the reinforced few-shot retrieval was fine-tuned with the reranker's training data.}
\label{tab:dev_final_score}
\end{table*}

\end{document}